\documentclass{llncs}

\usepackage[T1]{fontenc}

\usepackage{graphicx}
    \graphicspath{{figures/}}

\usepackage{color}

\usepackage{fancyvrb} 

\usepackage{hyperref}
    \hypersetup{
        pdftitle  = {Generic Library Interception for Improved Performance Measurement and Insight},                  
        pdfauthor = {Ronny Brendel, Bert Wesarg, Ronny Tsch\"uter, Matthias Weber, Thomas Ilsche and Sebastian Oeste} 
    }

\usepackage{url}

\usepackage{subfigure} 

\newcommand\footnoteWithoutMarker[1]{%
    \begingroup
    \renewcommand\thefootnote{}\footnote{#1}%
    \addtocounter{footnote}{-1}%
    \endgroup
}

\begin{document}

\title{Generic Library Interception for Improved Performance Measurement and Insight}

\author{Ronny~Brendel\inst{1} \and Bert~Wesarg\inst{2} \and Ronny~Tsch\"uter\inst{2} \and Matthias~Weber\inst{2} \and Thomas~Ilsche\inst{2} \and Sebastian~Oeste\inst{2}}

\institute{Oak Ridge National Laboratory, USA \\
    \email{brendelr@ornl.gov}
    \and
    Technische Universit\"at Dresden, Germany \\
    \email{\{bert.wesarg, ronny.tschueter, matthias.weber, thomas.ilsche, sebastian.oeste\}@tu-dresden.de}
}

\maketitle

\begin{abstract}

As applications grow in capability, they also grow in complexity.
This complexity in turn gets pushed into modules and libraries. 
In addition, hardware configurations become increasingly elaborate, too.
These two trends make understanding, debugging and analyzing the performance of applications more and more difficult.

To enable detailed insight into library usage of applications,
we present an approach and implementation in \mbox{Score-P} that supports intuitive and robust creation of wrappers for arbitrary C/C++ libraries.
Runtime analysis then uses these wrappers to keep track of how applications interact with libraries, how they interact with each other, and record the exact timing of their functions.
\footnoteWithoutMarker{%
This manuscript has been authored by UT-Battelle, LLC,
under contract DE-AC05-00OR22725 with the US Department of Energy (DOE).
The US government retains and the publisher, by accepting the article
for publication, acknowledges that the US government retains a
nonexclusive, paid-up, irrevocable, worldwide license to publish or
reproduce the published form of this manuscript, or allow others to do
so, for US government purposes. DOE will provide public access to these
results of federally sponsored research in accordance with the DOE
Public Access Plan (\url{http://energy.gov/downloads/doe-public-access-plan}).}

\end{abstract}

\section{Introduction}
\label{sec:intro}

\begin{figure}[tbh]
    \centering
    \includegraphics[width=0.9\columnwidth]{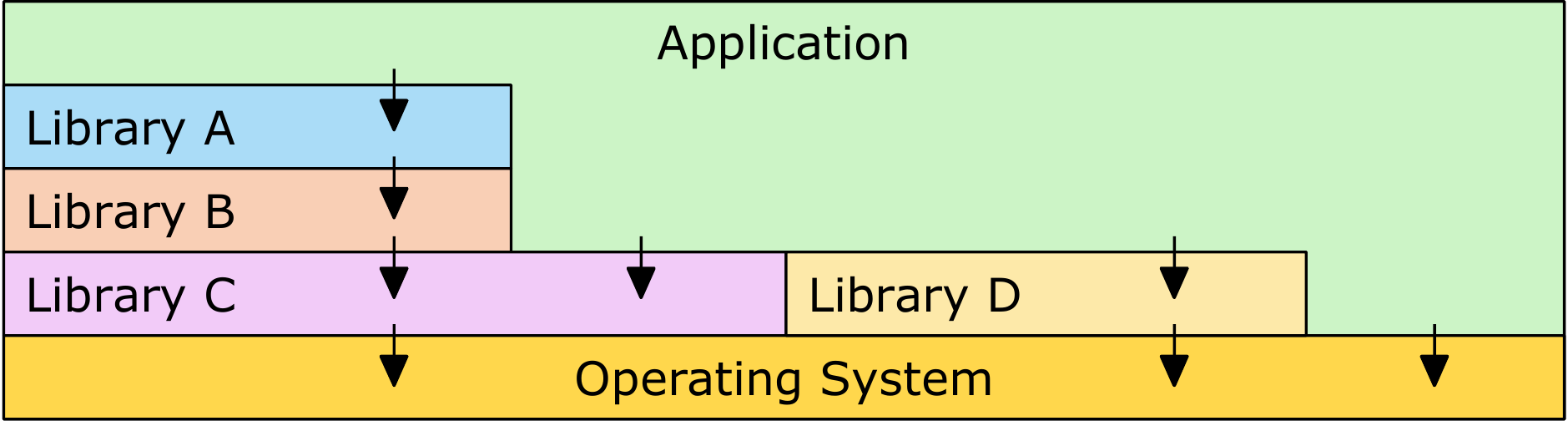}
    \caption{Typical software stack with an application relying on four libraries.}
    \label{fig:software_stack}
\end{figure}

To push science and businesses further, today's software becomes increasingly powerful but also complex.
Software libraries allow offloading this complexity into subunits, so that developers can focus on adding functionality by using them, rather than implementing every detail themselves.
But the complexity does not disappear --- It gets pushed down to lower levels.
The gained development convenience is traded for an increased effort of debugging and overall reasoning about the application including its performance characteristics.
Figure~\ref{fig:software_stack} depicts a typical application and its software dependencies.

A similar development takes place in computer architecture.
The adoption of multiple cores per CPU, heterogeneous architectures, complex cache/memory hierarchies, elaborate interconnect networks, as well as deep I/O hierarchies gives rise to a multitude of potential performance problems.
This increasing complexity in software and hardware makes performance analysis an integral part of the software life cycle.

Tool chains providing modern performance analysis capabilities include Linux Perf~\cite{DeMelo:2009}, NVIDIA profiling tools~\cite{NvidiaProfiling}, Intel~VTune~Amplifier~\cite{Intelvtune}, \mbox{Score-P}~\cite{Scorep:2012}, Arm~MAP~\cite{ArmMap} and HPCToolkit~\cite{Adhianto:2010}.
These tools combine multiple data collection techniques, like sampling, call stack unwinding, tools interfaces, library wrapping, compiler instrumentation, and manual instrumentation in various ways.
The goal is to gather data as detailed as needed while alleviating the disadvantages of individual techniques.
For example, it is common to combine sampling and call stack unwinding with library wrapping for important libraries.
Sampling gives coarse-grained stochastic timing information of the application's function call sequence, while library wrappers count and measure exact timings of library calls.
Aside from counting calls and measuring time, libraries like POSIX Threads~\cite{Butenhof:1997}, and I/O libraries like HDF5~\cite{Folk:2011} and ADIOS~\cite{Lofstead:2008} are commonly wrapped to extract semantic information (e.g. written bytes) from arguments passed to library functions.

Which wrappers are available is limited to what each performance tool supports.
Even if an application developer is interested in exact function call tracking of certain libraries, there is no well supported way to achieve this in any performance analysis tool today.

To address this, we introduce \emph{user library wrapping}, which is included in the upcoming release of the open-source performance monitor \mbox{Score-P}.
The feature empowers application developers to easily generate library wrappers for any C/C++ library.
This is significant, because:
\begin{itemize}
    \item With just link time changes, developers can now get exact performance information on any C/C++ library they want.
    \item They can analyze closed-source libraries, like the Intel MKL~\cite{Mkl}.
    \item They can track function calls from a library to itself and between libraries.
\end{itemize}

\mbox{Score-P} benefits from user library wrapping for the following reasons.
First, regular compiler instrumentation provides no call-backs upon library entry/exit.
Second, compiler instrumentation often yields high event rates, which leads to diminished performance and large event recordings.
This necessitates a filtering workflow that in turn complicates the whole measurement process.
With user library wrapping, developers can forego compiler instrumentation and still capture critical performance data, and have a small low-overhead recording at the same time --- no filtering needed.
Third, wrappers give exact function call counts and timings as opposed to the statistical information from sampling and call stack unwinding.
Additionally, it simplifies creating fixed wrappers that capture library semantics not only in \mbox{Score-P}, but for all tools that rely on library wrappers.

We took great care to make wrapper creation and usage as intuitive and simple as possible.
Numerous checks with polished error messages ensure the wrapper works correctly or let the developer know why it might not.

This paper is divided as follows:
Section~\ref{sec:related} enumerates related work.
Section~\ref{sec:method} first presents basics on library wrapping.
It then details the workflow for creating and using wrappers while highlighting some implementation choices,
by the example of wrapping the QtWidget and QtGui modules.
Section~\ref{sec:case} demonstrates how our approach aids investigating the performance characteristics of two real-world scientific applications.
The last two sections offer conclusions and indicate points of interest for future development.

\section{Related Work}
\label{sec:related}

Wrapping C/C++ libraries is not new.
SWIG~\cite{Beazley:1996}, first released in 1996, generates wrappers for C/C++ libraries so they can be called from other languages like Python, Go and Lua.
SWIG does not provide library interception for extracting, e.g., performance data.
Furthermore, it is not possible to create C++ or C wrappers for C/C++ libraries.
SWIG uses its own C/C++ preprocessor and parser.

Recently, Google released the C++ Language Interface Foundation (CLIF)~\cite{Clif} which provides similar functionality to SWIG.
It uses Clang~\cite{Clang} to analyze the library headers, and for now only generates wrappers for Python.

Some libraries, like OpenMP~5~\cite{Eichenberger:2013} and CUDA~\cite{Cuda}, offer a so-called \emph{tools interface} for analysis tools to hook into.
In that case, library wrapping is not needed.
But for most libraries, wrappers are required to gain insight into their usage.
MPI~\cite{MPI} provides a special \emph{profiling interface}, which helps create wrappers by providing all functions as weak symbols so that they can be overridden.
Wrapper functions call the actual MPI functions through a \texttt{P}-prefixed symbol, e.g. \verb+PMPI_Send+.

One possible application for tools interfaces and library wrappers is to check for correct API usage.
For example MUST~\cite{Hilbrich:2010} uses MPI's profiling interface to ensure correct use and to detect possible deadlocks.
The wrapping code is generated manually with a simple proprietary wrapper generator.

Software performance analysis tools commonly use fixed wrappers to gain insight into the use of specific libraries.
For example Arm~MAP~\cite{ArmMap} is a commercial profiler specialized in analyzing multi-paradigm applications.
It wraps MPI and OpenMP functions and uses the tools interface of CUDA.
Various open-source performance analysis tools exist.
Some of them are Extrae~\cite{Extrae}, HPCToolkit~\cite{Adhianto:2010} and \mbox{Score-P}~\cite{Scorep:2012}.
All three support a variety of parallelization schemes and hardware platforms.
They differ in techniques, focus and user interface, but are similar in terms of utilizing library wrapping.

VampirTrace includes a simple implementation of user library wrapping~\cite{Dietrich:2010}.
It is based on CTool~\cite{Ctool} (abandoned in 2004), supports only C, has several technical limitations and needs manual intervention in most cases.
\mbox{Score-P} is VampirTrace's successor.

TAU offers user library wrapping via the \texttt{tau\_wrap} and \texttt{tau\_gen\_wrapper} commands~\cite{Shende:2011}.
It uses the Edison~Group's commercial C/C++ parser~\cite{Edg}.
TAU's implementation has multiple limitations. For example, it does not support C++, cannot wrap functions with function pointers or ellipsis arguments, and compile and link flags are not customizable.


\section{Methodology}
\label{sec:method}

Our goal is to provide a simple and robust way to record performance data on library function calls.
For this, we need an opportunity to intercept them.
That means whenever a library function is called, the measurement system has to be invoked.

\subsection{Library Call Interception}
\label{sec:method:library}

We distinguish two wrapping methods based on when interception is set up: \emph{link time} and \emph{runtime}.
These two methods also differ in the kind of functions that can be intercepted.

\textbf{Link time}:
The first approach is based on the \verb+--wrap+ option of the GNU linker\footnote{\url{https://sourceware.org/binutils/docs-2.28/ld/Options.html}}.
For example, to wrap the function \texttt{foo}, we have to implement the corresponding wrapper function \verb+__wrap_foo+.
The original function is available via the \verb+__real_foo+ symbol.
Then specifying \verb+--wrap foo+ in the link command enables wrapping \texttt{foo}, and the GNU linker resolves these symbols appropriately.
This approach is limited to instances where the link step of the application can be modified, as the symbols of interest need to be specified at link time.
Wrapping symbols called from shared libraries does not work, because the linker resolves these symbols at runtime.

\textbf{Runtime}:
At the start of executing an application, the dynamic linker loads and links all dependent shared libraries.
The second approach modifies the order in which the dynamic linker loads them.
To wrap a function, we provide a replacement function with the same symbol name as the wrapped function.
The linker, then, needs to link the wrapper before the target library.
One way to achieve this is modifying the link step to put the wrapper library before the original one.
Alternatively, let the environment variable \verb+LD_PRELOAD+\footnote{\url{http://man7.org/linux/man-pages/man8/ld.so.8.html}} point to the wrapper library before executing the application.
The latter method has the advantage that it does not need to modify the link step.
Once called, the wrapper function loads the target library via \texttt{dlopen},
searches for the address of the original symbol using \texttt{dlsym},
and then forwards the original call.
With link step modification, this approach can intercept all calls that link time wrapping can, plus those that originate from shared libraries.
The \verb+LD_PRELOAD+-based version can only intercept calls to shared libraries, not to statically linked ones.

Both mechanisms require wrappers that pose as the original functions.
For each call, a wrapper function notifies the performance monitor before and after forwarding the call to the original function.
In the next section we present the workflow with which users create their own library wrapper.

\subsection{Workflow}
\label{sec:method:work}

\begin{figure}[tbh]
    \centering
    \includegraphics[width=0.9\columnwidth]{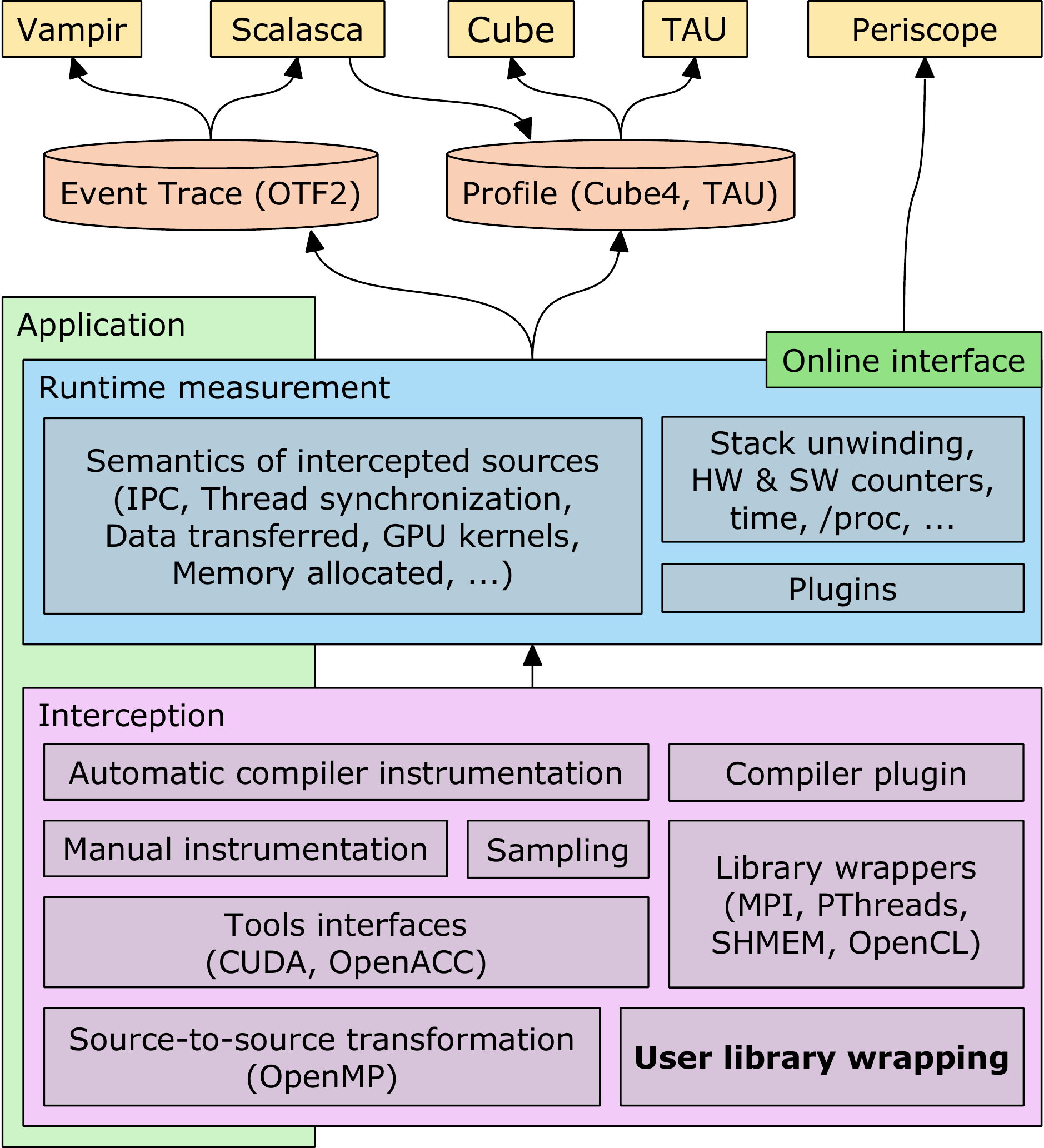}
    \caption{Overview of the \mbox{Score-P} measurement system architecture. User library wrapping provides an additional interception mechanism.}
    \label{fig:scorep_architecture}
\end{figure}

In this work we extend \mbox{Score-P} --- a state-of-the-art software performance monitor.
Figure~\ref{fig:scorep_architecture} shows its high-level architecture.

The goal is to make calls to library functions available for performance analysis.
For this, we add functionality to record timestamped \emph{enter}- and \emph{exit}-events for these calls.

\begin{figure}[tbh]
    \centering
    \includegraphics[width=0.9\columnwidth]{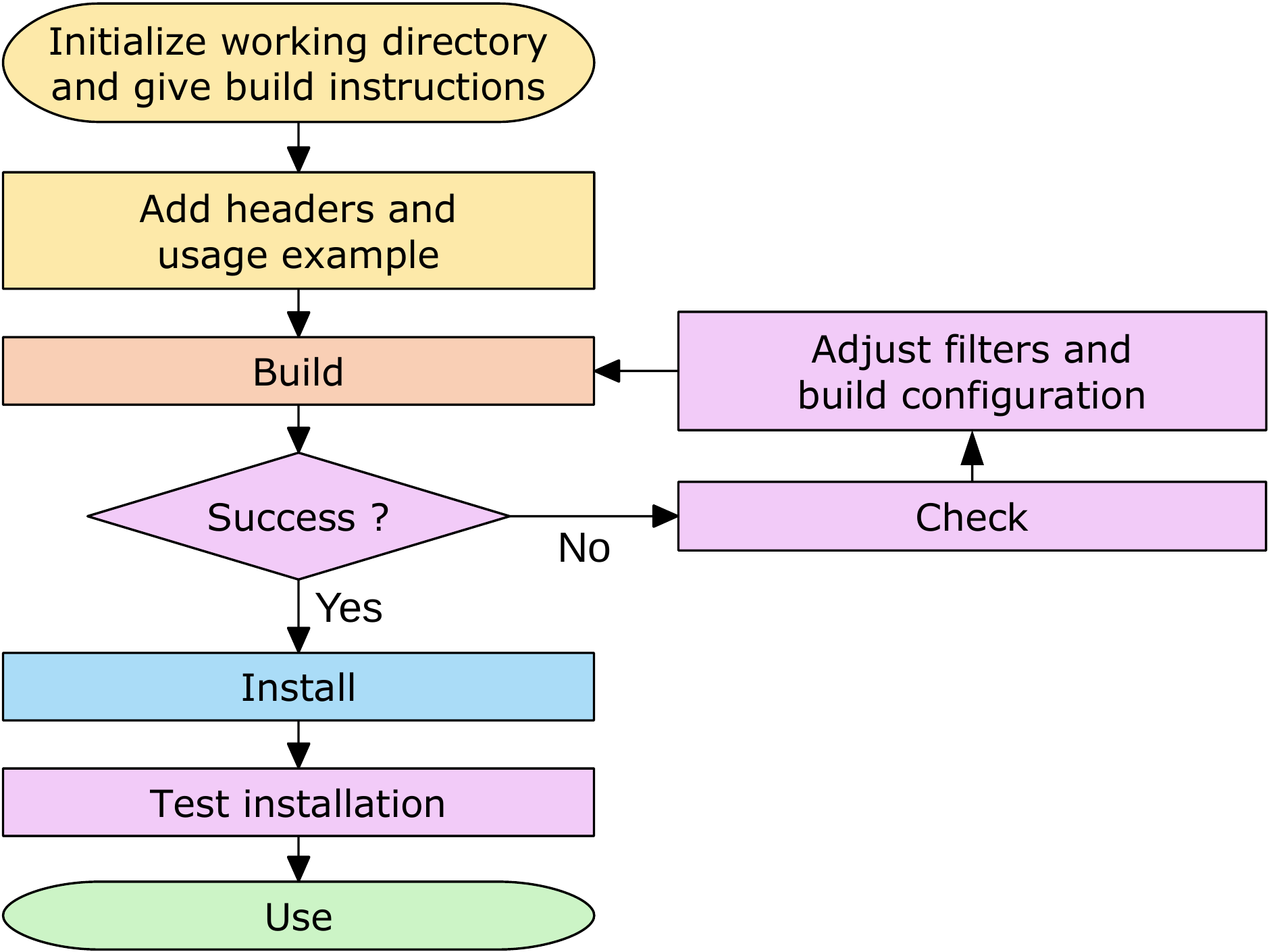}
    \caption{High-level workflow for creating a user library wrapper.}
    \label{fig:workflow_overview}
\end{figure}

The process of generating a library wrapper is intricate and error-prone.
Thus, the highest priority in the design of user library wrapping is to make it reliable.
To guide the user through these potential problems we introduce a workflow,
which the following paragraphs explain.
We motivate some of the implementation choices by highlighting the intricacies that necessitate them.
Figure~\ref{fig:workflow_overview} depicts the steps involved.

\subsubsection{Initialize the Working Directory}
\label{sec:method:work:init}

The tool \texttt{scorep-libwrap-init} initiates the bootstrapping process.
For this it creates a working directory where all subsequent steps take place in.
The command takes a number of arguments that concern compilation setup, linking setup, and the name of the user library wrapper.
Essentially, the user specifies how to compile and link an application using the target library.

In this step, \mbox{Score-P} tries to locate potential shared versions of the target libraries.
It lets the user know if it cannot find any to avoid confusion due to failing \texttt{dlopen}-calls later on.

\texttt{scorep-libwrap-init} creates a number of files in the new working directory:
\begin{itemize}
    \item A detailed documentation with explanations of possible warnings and errors
    \item A \texttt{Makefile} that guides the user through the next steps
    \item Stub source-, header-, and filter-files, which subsequent paragraphs explain
\end{itemize}
At the end, the command prints out what the next steps are.

For example, the following command initializes the working directory for a library wrapper of the QtGui and QtWidget modules~\cite{Qt}%
\footnote{Full example: \url{https://github.com/score-p/scorep_libwrap_examples/tree/1564c272311d04575f9886cd982fc611e07eb295/qt5/qtgui-and-qtwidgets}}%
:
\begin{Verbatim}[samepage=true]
    $ scorep-libwrap-init -x c++                 \
        --name qtgui_and_qtwidgets               \
        --display-name "Qt Gui & Widgets"        \
        --cppflags "-fPIC -I${QT_INCLUDE}"       \
        --ldflags "-fPIC"                        \
        --libs "-lQt5Widgets -lQt5Gui -lQt5Core"
\end{Verbatim}

\subsubsection{Add Library Headers}
\label{sec:method:work:headers}

Next, the user adds an include-statement for each header an application usually includes from the target library to \texttt{libwrap.h}.
This approach allows the user to specify a sequence of includes and preprocessor macros,
which the wrapper generator can then process.
Continuing the example, we add:
\begin{Verbatim}[samepage=true]
    #include <QtGui/QtGui>
    #include <QtWidgets/QtWidgets>
\end{Verbatim}

\subsubsection{Create an Example Application}
\label{sec:method:work:example}

To be able to verify the results, the process needs a test case.
For this, the user adds a small usage example to \texttt{main.c}/\texttt{cc}.
It will be compiled, linked and executed later to test whether the target library and wrapper work.
Continuing the example, we write a simple Qt application that opens a window and creates an unused image:
\begin{Verbatim}[samepage=true]
    int main(int argc, char** args) {
        QApplication app(argc, args);
        QWidget w;
        QImage i;
        w.show();
        return app.exec();
    }
\end{Verbatim}

\subsubsection{Create the Wrapper}
\label{sec:method:work:make}

Before building the wrapper, the user can always adjust the compile- and link-setup by either directly changing the top lines in the makefile,
or by invoking \texttt{scorep-libwrap-init} again with the \verb+--update+ argument.

The user can now attempt to build the wrapper via the \texttt{make}-command.
First, this links the example application to the target library.
If that fails, the provided example is wrong.

Next, it preprocesses \texttt{libwrap.h} to create \texttt{libwrap.i} with the same compiler and flags that are used to create the provided example.
Our \emph{libclang}-based analyzer then processes this file to generate the complete list of library functions (plus name spaces, classes and types).
During this step, the analyzer consults a \emph{filter file} for functions to ignore.

The generated list of functions is then used to create an example application, which contains a call to each of these functions.
If linking this application to the target library fails, there are wrapper functions that do not have an original function in the target library.
For example this happens for some class constructors and inline functions.

If \texttt{make} fails, the next step is \texttt{make check}, which the next paragraph explains.
If \texttt{make} passes, it creates the wrapper.
The wrapper consists of up to four different wrapper libraries.
One dimension is whether the wrapper is a shared or static library.
The other dimension is whether the wrapper contains the code for link time or runtime wrapping.
All four versions are useful depending on the application/library/system setup.
If this succeeds, the user can move on and install the wrapper.

While processing the header files, there are a number of warnings and errors that can occur.
For example, the wrapper warns about functions that contain ellipsis arguments, because they cannot be forwarded in C.
In case a \texttt{v}-version (like \texttt{vprintf} is to \texttt{printf}) exists, the \verb+LIBWRAP_ELLIPSIS_MAPPING_SYMBOLS+-variable in the makefile
lets the user create a mapping so the wrapper can forward the call to the \texttt{v}-version via the \verb+va_arg+ argument.

In C, having an empty argument list in a function declaration means the argument list is unknown, i.e. \emph{variadic}.
In C++, on the other hand, the same syntax means it is really empty.
In C, you need to use \texttt{(void)} as argument list for this.
Calling a variadic function without parameters is valid C.
This means valid C can trip up library wrapping, if the library developer did not use \texttt{(void)} for an empty argument list.
To work around this, the makefile provides the \verb+LIBWRAP_VARIADIC_IS_VOID_SYMBOLS+-variable,
which names functions that are to be treated as having an empty argument list.

In our example, the \texttt{make}-step warns about a number of ellipsis functions, for example in the \texttt{QMessageLogger} and \texttt{QString} classes.
It exits with an error message, because there is a mismatch between functions found in the headers and the symbols in the library.
To find out which functions these are, we need to run \texttt{make check} and then adjust the filter.

\subsubsection{Verify the Wrapper}
\label{sec:method:work:make-check}

Because the function list generated by the library header analysis rarely matches the symbol table in the library,
for each wrapper function, \texttt{make check} generates a source file, and tries to compile and link it with and without the target library.
The result is a complete list of symbols that are missing from the target library,
and a list of symbols where linking works even without the target library.
The latter tries to weed out functions that are not intended to be wrapped, because they are in system libraries.%
\footnote{Creating a source file for each function and try compiling and linking it is a common technique among configure tools.
Doing this in one compile-link-step would require parsing the output of each supported compiler and version,
which is not portable across compilers and linkers.}

Using the two generated lists, the user has to adjust the filter to remove unwrappable, and perhaps some unwanted functions.
This not only ensures the soundness of the wrapper,
but also makes sure the user chooses the functions deliberately.
Accidentally wrapping more than intended should be avoided.

After this, the user has to repeat \texttt{make} and \texttt{make check} until \texttt{make} succeeds.

Executing this step in our example first informs us that it is doing this check for over $13,000$ functions, and this may take some time.
Looking at the list of these functions (in the \texttt{.wrap}-file), we notice that it wraps more than just QtGui and QtWidget's components.
This is because the header analysis cannot read the users intention perfectly.
It initially only includes functions that it finds in files in directories specified via the \texttt{-I}-compilation-flag.%
\footnote{Not doing this would initially always wrap everything including functions from system headers.}
Thus we refine the filter from
\begin{verbatim}
    INCLUDE /usr/include/x86_64-linux-gnu/qt5/*
\end{verbatim}
to
\begin{Verbatim}[samepage=true]
    INCLUDE /usr/include/x86_64-linux-gnu/qt5/QtGui/*
    INCLUDE /usr/include/x86_64-linux-gnu/qt5/QtWidgets/*
\end{Verbatim}
and repeat \texttt{make check}.
This yields a list of 818 missing functions, which we add to the filter.
No symbols were found that exist when not linking to Qt.

Repeating \texttt{make} still fails due to a restriction in libclang with C++.
If a function uses a type that is created via \texttt{typedef} or \texttt{using} in a class, our header analysis cannot always determine the fully qualified type.
This case requires user intervention.
In our example we can fix this by looking up the types in Qt's documentation and adding the class scopes via text replacement to the wrapper code.%
\footnote{\url{https://github.com/score-p/scorep_libwrap_examples/blob/1564c272311d04575f9886cd982fc611e07eb295/qt5/qtgui-and-qtwidgets/fix-type-scopes.sh}}

\subsubsection{Install the Wrapper}
\label{sec:method:work:make-install}

Once the wrapper builds, \texttt{make install} installs it.
If not specified otherwise, this installs the wrapper into \mbox{Score-P's} installation directory.

\subsubsection{Verify the Installed Wrapper}
\label{sec:method:work:make-installcheck}

Invoking \texttt{make installcheck} links the example application to the link time and runtime wrapper library in the same way the user would.
This step creates two executables, and prints out how to run and check the resulting \mbox{Score-P} measurement.

\begin{figure}[tbh]
    \centering
    \includegraphics[width=0.9\columnwidth]{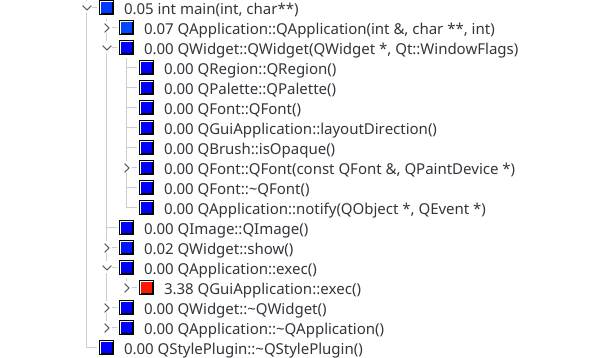}
    \caption{Partially collapsed Cube profile of the Qt example application. It accurately resembles the source code. Numbers are seconds, inclusive execution time for collapsed entries and exclusive for expanded ones.}
    \label{fig:qt_libwrap_profile}
\end{figure}

Running the example yields a profile with over 5000 calls to 251 unique Qt functions.
Figure~\ref{fig:qt_libwrap_profile} shows an excerpt.
Without the wrapper, compiler instrumentation would only recognize and record the \texttt{main}-function.
Sampling with stack unwinding yields a more detailed call graph (e.g., it includes system and desktop system functions),
but misses many function calls due to the nature of sampling, and it also cannot capture exact timing and call counts.

\subsubsection{Use the Wrapper}
\label{sec:method:work:use}

If the wrapper has not been installed into \mbox{Score-P's} installation directory (the default),
the environment variable \verb+SCOREP_LIBWRAP_PATH+ (\verb+PATH+-like) needs to point to the wrapper's path before using it.

\mbox{Score-P's} new \verb+--libwrap=<wrappername>+-flag then modifies the link step to activate one or more wrappers.

To use our example Qt wrapper,
we link the to-be-analyzed application according to the instructions we initially gave \texttt{scorep-libwrap-init}
and simply prefix it with \verb+scorep --libwrap=qtgui_and_qtwidgets+. I.e.:
\begin{Verbatim}[samepage=true]
    $ scorep --libwrap=qtgui_and_qtwidgets g++ \
        -fPIC -I${QT_INCLUDE}                  \
        application.cc                         \
        -fPIC -lQt5Widgets -lQt5Gui -lQt5Core  \
        -o application
\end{Verbatim}

Optionally, the user can specify the wrap method by prefixing the wrapper name with either \texttt{linktime:} or \texttt{runtime:}.

\subsubsection{Auxiliary Commands}
\label{sec:method:work:aux}

In HPC centers, we expect support staff, not only users themselves, to install wrappers of analysis-worthy libraries alongside a \mbox{Score-P} installation.
Users can still install wrappers into their own directories.
One advantage is that the staff can update the wrappers at the same time as they update \mbox{Score-P} or the target libraries.

The command \verb+scorep --help+, among other information, gives a list of installed wrappers.
Users can invoke \texttt{scorep-info libwrap-summary}, with an optional wrapper name, to view wrapper configurations in greater detail.

\subsection{Implementation Details}
\label{sec:method:implementation}

Because compile-time commands, e.g., \verb+#ifdef+, can influence the list of declared functions,
we decided to employ the user's compiler to preprocess the library's headers.
To generate this list of functions, we read the header using libclang.
This mismatch between preprocessor and reader can sometimes lead to problems because they might not agree on the language standard to use.
Specifying the standard explicitly solves this.

During development we realized that \mbox{Score-P's} configured compiler cannot always link libclang to the wrapper generator.
The configure step would need to know the compiler with which libclang has been created.
To circumvent this, contrary to other parts of \mbox{Score-P}, it builds the wrapper generator using Clang, if available.

The presented approach relies on wrapping facilities offered by the linker and dynamic linker.
Many C++ libraries heavily rely on inlining and templates.
Wrapping libraries based on symbols being present in the target library means that this technique is unable to intercept inlined function calls.

\section{Case Study}
\label{sec:case}

The previous section proves that our approach is robust by wrapping two Qt modules.
This section demonstrates how user library wrapping benefits performance analysis for two real-world scientific applications.
We repeat all measurements five times, and pick the median.

\subsection{GROMACS}
\label{sec:case:gromacs}

GROMACS~\cite{Vanderspoel:2005} is a popular molecular dynamics package specialized in simulating proteins, lipids and nucleic acids.
To leverage the compute power of HPC sytems, GROMACS relies on MPI, OpenMP, CUDA and either FFTW~3~\cite{Frigo:1998} or the Intel Math Kernel Library for discrete Fourier transforms.

For our demonstration we use GROMACS' current version $2016.3$, and simulate a lysozyme in water~\cite{Lemkul} using one tenth of the default number of time steps.
We run the simulation on Oak Ridge National Laboratory's Titan, a Cray XK7 supercomputer.
Each node has one AMD Opteron~6274 CPU with eight Bulldozer modules and one NVIDIA~Tesla~K20X graphics card.
We choose to run on two nodes, with four processes each. Every process spawns one additional thread --- a total of 16 threads.
On the software side, we load the default GNU-based environment, which uses GCC~4.9.3 and FFTW~3.3.4.11.

Executing GROMACS normally takes 330~seconds, of which it spends 193 in the main part, the actual simulation of the protein (\emph{Production MD}).

To instrument GROMACS with \mbox{Score-P}, we replace the compilers \texttt{cc} and \texttt{CC} in the CMake-command with \mbox{Score-P's} compiler wrappers \texttt{scorep-cc} and \texttt{scorep-CC} and prefix the command with \verb+SCOREP_WRAPPER=off+.
Building works the same as before.
\mbox{Score-P} then, by default, enables automatic compiler instrumentation and injects the performance monitor by modifying compile and link commands.
We only use this instrumented GROMACS build on the expensive Production MD part, and execute all other parts with the normal build.
Executing this increases Production MD's execution time to 375~s (+94.3\%).
\mbox{Score-P} registers 3.04~billion function calls, 2.96~billion of which are user functions.
The other 80~million are OpenMP loops/calls and MPI calls.
\texttt{scorep-score} estimates that a trace of this execution is 76~gigabytes (GB) large.

For technical reasons \mbox{Score-P} requires instrumenting MPI and OpenMP events.
Therefore, a reduced recording without any user functions takes 214~s (+10.9\%), contains 79~million calls, and a trace of this configuration is 3.0~GB large.

\mbox{Score-P's} default (automatic compiler instrumentation) adds significant overhead, and should not be used in tracing mode as is.
By following \mbox{Score-P's} filtering workflow we can reduce the overhead and trace recording size.
Alternatively we can switch off compiler instrumentation to record a very small amount of information.
But none of these three options record anything about FFTW.

To track calls into FFTW, we need to create a wrapper library for it following the workflow described in Section~\ref{sec:method:work}.
One thing that confuses our process is that Cray's compiler wrapper \texttt{cc} pulls in modules, like FFTW, automatically, if the module is loaded.
Thus, to compile a program using FFTW we don't need to add compile and link flags.
This is not a problem, but disarms one of our checks and makes wrapper creation slightly confusing.
To circumvent this we change the environment variable \verb+PE_PKGCONFIG_PRODUCTS+ to not include \verb+PE_FFTW+.
The full instructions for building the wrapper are available online\footnote{\url{https://github.com/score-p/scorep_libwrap_examples/tree/1564c272311d04575f9886cd982fc611e07eb295/fftw3}}.

To configure GROMACS with the FFTW wrapper, run CMake with \mbox{Score-P's} compiler wrappers as in the previous case.
Then build it using
\begin{verbatim}
    SCOREP_WRAPPER_INSTRUMENTER_FLAGS="--libwrap=fftw3" make
\end{verbatim}
instead of just \texttt{make} to enable the wrapper.
With this, Production MD takes 214~s (the same as the minimum instrumentation) and counts additional 5.9~million function calls.
The corresponding trace is 3.1~GB large.

\begin{figure*}[tbh]
  \centering
  \subfigure[Top: Overview of the whole execution. Bottom left: Accumulated (over all threads and processes) exclusive time for all FFTW functions. Bottom right: Accumulated invocation counts.]{
    \label{fig:gromacs-1}
    \includegraphics[width=0.90\linewidth]{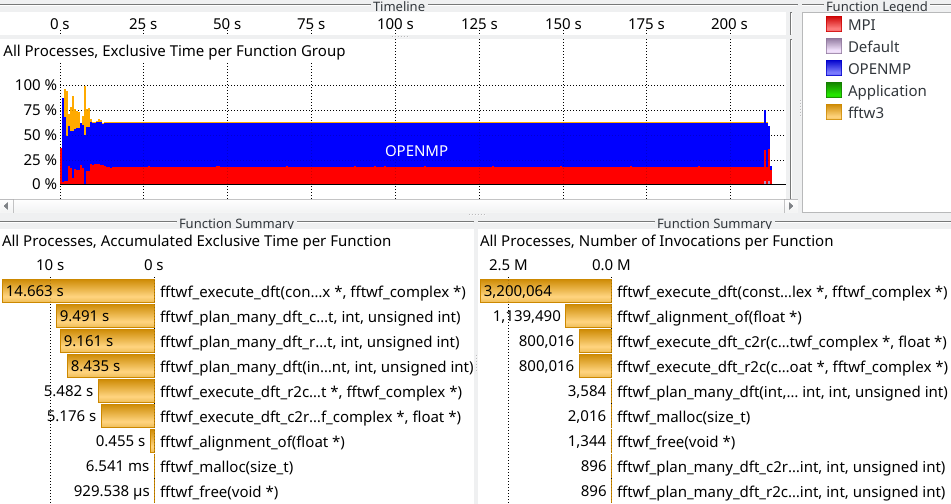}
  }
  \subfigure[Top: Currently active function active per process and thread. Orange is FFTW. Bottom: FLOPS encoded in color displayed in the same layout.]{
    \label{fig:gromacs-2}
    \includegraphics[width=0.90\linewidth]{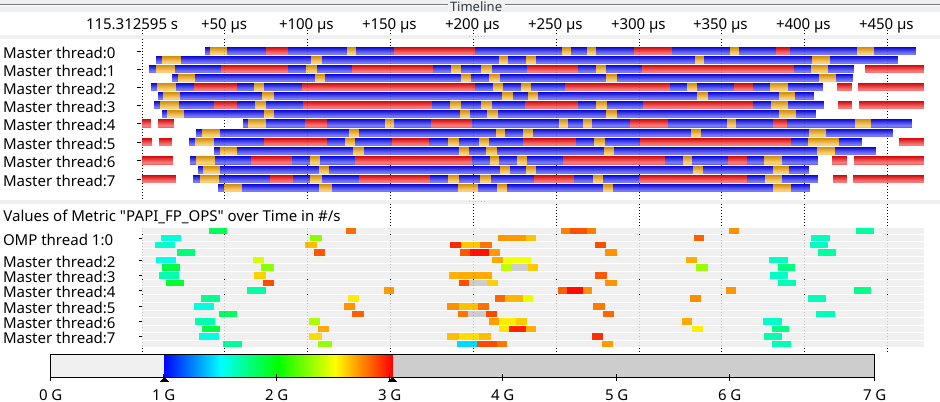}
  }
  \caption{Visual analysis of the trace run of GROMACS with Vampir~\cite{Brunst:2013}}
  \label{fig:gromacs}
\end{figure*}

By analyzing this recording, we discover that GROMACS spends the majority of time in OpenMP loops (Figure~\ref{fig:gromacs-1}).
FFTW occupies only about 2.4\% of the execution time.
Nevertheless, for the low amount of time spent in it, there are a lot of calls to FFTW.
This suggests that, if possible, putting more work into one iteration should be considered.
Because the vast majority of calls to FFTW take below three milliseconds, a sampling-based analysis would show a distorted picture.

With this exact instrumentation of FFTW, we can now, for example, investigate how efficiently it exploits the underlying hardware by recording performance counters.
Figure~\ref{fig:gromacs-2} shows how FFTW's use of the floating-point unit varies between calls, and is generally subpar.

\subsection{PERMON}
\label{sec:case:permon}

The software package PERMON~\cite{Hapla:2015} solves quadratic programming problems with the help of FETI~methods~\cite{Farhat:1991} for domain decomposition.
PERMON extends PETSc~\cite{Balay:2016} and is used mainly for simulating mechanical structure, for example linear elasticity, elasto-plasticity and shape optimization.

\mbox{Score-P} offers multiple ways to analyze the interplay of PERMON and PETSc.
One is to instrument both.
A second way is to analyze only PERMON and use sampling and call stack unwinding to peek into PETSc.
A third approach is to intercept all function calls to PETSc by creating a wrapper for it.
\mbox{Score-P} also supports combinations of instrumentation, sampling with call stack unwinding and library wrapping.

Additionally instrumenting PETSc is cumbersome and means creating a custom installation just for measurement with \mbox{Score-P}.
The second way is good from an ease-of-use perspective, but has drawbacks.
It does not record all PETSc calls, cannot count the number of calls, and cannot record the exact timing of calls.
Employing user library wrapping yields a good level of detail while alleviating the drawbacks of the other two methods.

\begin{figure}[tbh]
    \centering
    \includegraphics[width=0.9\columnwidth]{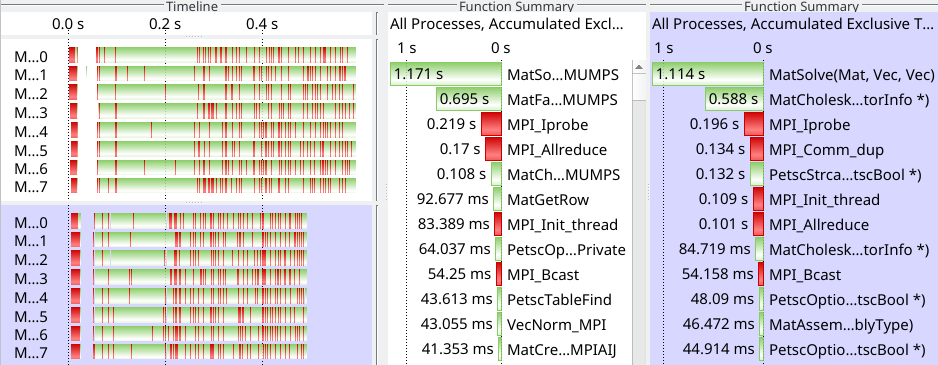}
    \caption{Vampir Master Timeline and profile excerpt of a PERMON run using eight MPI processes. White background: instrumenting both PERMON and PETSc, blue background: user library wrapping}
    \label{fig:permon}
\end{figure}

Figure~\ref{fig:permon} shows the resulting traces from the first and third approach side-by-side.
The full instrumentation creates a 112 megabyte trace,
whereas the run without automatic compiler instrumentation and with library wrapping results in a 84~megabyte recording.
Both recordings are similar in detail and characteristics.

\pagebreak
\section{Conclusions}
\label{sec:conclusions}

In this work we present \emph{user library wrapping}, an extension to \mbox{Score-P} that allows exact tracking of function calls to any C/C++ library.
It enables in-depth performance analysis of applications in conjunction with their underlying libraries.
Furthermore it tracks calls between libraries and offers insight into closed-source libraries like the Intel Math Kernel Library.

We offer a simple, well-crafted workflow to create and use library wrappers.
This workflow guides the user through an otherwise difficult procedure, and minimizes mistakes.

Our approach differs from previous incarnations in that it supports C++, is mature, robust and well documented.
It requires minimal manual work and uses modern Clang/LLVM facilities to analyze library headers.

We demonstrate its robustness for non-trivial use-cases by wrapping the QtWidgets and QtGui modules.
Furthermore we show how user library wrapping enables better performance analysis for two real-world scientific applications.

\section{Future Work}
\label{sec:future}

There are multiple interesting areas to pursue.
By using compilation databases provided by CMake and GNU Autotools, we might be able to drop the requirement to specify how to build an example application in the first workflow step.

Because HPC systems install multiple versions of the same library, it would be beneficial to explicitly support versioning.

The presented approach forwards parameters from the wrapper to the target function, but does nothing with it.
Extending our approach to record parameter values, for example like a performance counter, can be useful.

In order to ensure the soundness of each wrapper, the presented workflow involves a number manual checks.
Technically, each wrapper needs to updated if its target library or Score-P is updated.
Repeating this procedure for every update is unnecessarily burdensome.
Therefore, the workflow should be extended to include automatic updating of generated wrappers.

Due to close consideration of the circumstances of header preprocessing and the symbol tables of library files,
each wrapper is tied to the machine it has been created on.
It should be investigated how to enable reusing the wrappers across different machines.
Ultimately, a public archive of wrappers is desirable.

\mbox{Score-P} hinges on modifying compile and link commands to instrument an application.
Some features, user library wrapping included, can be used with just link time changes.
But that is not strictly necessary.
By loading all of \mbox{Score-P} at runtime using \verb+LD_PRELOAD+, we could skip link command line changes,
and attach the performance monitor to an unmodified binary.

Because there are many runtime analysis tools relying on library wrappers,
we would like to offer our wrapper creation facility to these projects.
Up until now, there was no well-supported, generic way to wrap C/C++ libraries for analysis.
Developers need to create wrappers, regularly update them, and keep track of new versions of the target library.

\section*{Acknowledgments}
This research used resources of the Oak Ridge Leadership Computing Facility at Oak Ridge National Laboratory, which is supported by the Office of Science of the Department of Energy under Contract DE-AC05-00OR22725.

This work is supported in part by the German Research Foundation (DFG) within the CRC 912 - HAEC.

\bibliographystyle{splncs03}
\bibliography{paper-references}

\begin{thebibliography}{10}
\providecommand{\url}[1]{\texttt{#1}}
\providecommand{\urlprefix}{URL }

\bibitem{Adhianto:2010}
Adhianto, L., Banerjee, S., Fagan, M., Krentel, M., Marin, G., Mellor-Crummey,
  J., Tallent, N.R.: {HPCToolkit}: Tools for performance analysis of optimized
  parallel programs. Concurrency and Computation: Practice and Experience
  22(6),  685--701 (2010)

\bibitem{ArmMap}
Arm map --- arm (Aug 2017),
  \url{https://www.arm.com/products/development-tools/hpc-tools/cross-platform/forge/map}

\bibitem{Balay:2016}
Balay, S., Abhyankar, S., Adams, M., Brune, P., Buschelman, K., Dalcin, L.,
  Gropp, W., Smith, B., Karpeyev, D., Kaushik, D., et~al.: {PETSc} users manual
  revision 3.7. Tech. rep., Argonne National Lab.(ANL), Argonne, IL (United
  States) (2016)

\bibitem{Beazley:1996}
Beazley, D.M., et~al.: {SWIG}: An easy to use tool for integrating scripting
  languages with {C} and {C}++. In: Tcl/Tk Workshop (1996)

\bibitem{Brunst:2013}
Brunst, H., Weber, M.: Custom hot spot analysis of {HPC} software with the
  {V}ampir performance tool suite. In: Tools for High Performance Computing
  2012, pp. 95--114. Springer (2013)

\bibitem{Butenhof:1997}
Butenhof, D.R.: Programming with {POSIX} threads. Addison-Wesley Professional
  (1997)

\bibitem{Clang}
clang: a {C} language family frontend for {LLVM} (Aug 2017),
  \url{http://clang.llvm.org}

\bibitem{Clif}
google/clif: Wrapper generator foundation to wrap {C++} for {P}ython and other
  languages using {LLVM} (Aug 2017), \url{https://github.com/google/clif}

\bibitem{Ctool}
{CTool} library (Aug 2017), \url{http://ctool.sourceforge.net}

\bibitem{Cuda}
{CUDA} zone | {NVIDIA} developer (Aug 2017),
  \url{https://developer.nvidia.com/cuda-zone}

\bibitem{Dietrich:2010}
Dietrich, R., Ilsche, T., Juckeland, G.: Non-intrusive performance analysis of
  parallel hardware accelerated applications on hybrid architectures. In:
  Parallel Processing Workshops (ICPPW), 2010 39th International Conference on.
  pp. 135--143. IEEE (2010)

\bibitem{Edg}
Edison design group (Aug 2017), \url{http://edg.com}

\bibitem{Eichenberger:2013}
Eichenberger, A.E., Mellor-Crummey, J., Schulz, M., Wong, M., Copty, N.,
  Dietrich, R., Liu, X., Loh, E., Lorenz, D.: {OMPT}: An {OpenMP} tools
  application programming interface for performance analysis. In: International
  Workshop on OpenMP. pp. 171--185. Springer (2013)

\bibitem{Extrae}
{Extrae} $\mid$ {BSC} tools (Aug 2017), \url{https://tools.bsc.es/extrae}

\bibitem{Farhat:1991}
Farhat, C., Roux, F.X.: A method of finite element tearing and interconnecting
  and its parallel solution algorithm. International Journal for Numerical
  Methods in Engineering  32(6),  1205--1227 (1991)

\bibitem{Folk:2011}
Folk, M., Heber, G., Koziol, Q., Pourmal, E., Robinson, D.: An overview of the
  {HDF5} technology suite and its applications. In: Proceedings of the
  EDBT/ICDT 2011 Workshop on Array Databases. pp. 36--47. ACM (2011)

\bibitem{Frigo:1998}
Frigo, M., Johnson, S.G.: Fftw: An adaptive software architecture for the fft.
  In: Acoustics, Speech and Signal Processing, 1998. Proceedings of the 1998
  IEEE International Conference on. vol.~3, pp. 1381--1384. IEEE (1998)

\bibitem{Hapla:2015}
Hapla, V., Horak, D., Pospisil, L., Cermak, M., Vasatova, A., Sojka, R.:
  Solving contact mechanics problems with {PERMON}. In: International
  Conference on High Performance Computing in Science and Engineering. pp.
  101--115. Springer (2015)

\bibitem{Hilbrich:2010}
Hilbrich, T., Schulz, M., de~Supinski, B.R., M{\"u}ller, M.S.: {MUST}: A
  scalable approach to runtime error detection in {MPI} programs. In: Tools for
  High Performance Computing 2009, pp. 53--66. Springer (2010)

\bibitem{Scorep:2012}
Kn{\"u}pfer, A., R{\"o}ssel, C., an~Mey, D., Biersdorff, S., Diethelm, K.,
  Eschweiler, D., Geimer, M., Gerndt, M., Lorenz, D., Malony, A., et~al.:
  {Score-P}: A joint performance measurement run-time infrastructure for
  {Periscope}, {Scalasca}, {TAU}, and {Vampir}. In: Tools for High Performance
  Computing 2011, pp. 79--91. Springer (2012)

\bibitem{Lemkul}
Lemkul, J.A.: Gromacs tutorial: Lysozyme in water (Sep 2017),
  \url{http://www.bevanlab.biochem.vt.edu/Pages/Personal/justin/gmx-tutorials/lysozyme/index.html}

\bibitem{Lofstead:2008}
Lofstead, J.F., Klasky, S., Schwan, K., Podhorszki, N., Jin, C.: Flexible io
  and integration for scientific codes through the adaptable io system
  ({ADIOS}). In: Proceedings of the 6th international workshop on Challenges of
  large applications in distributed environments. pp. 15--24. ACM (2008)

\bibitem{DeMelo:2009}
de~Melo, A.C.: Performance counters on linux. In: Linux Plumbers Conference
  (2009)

\bibitem{Mkl}
Intel\textregistered~math kernel library (intel\textregistered~mkl) |
  intel\textregistered~software (Aug 2017),
  \url{https://software.intel.com/en-us/mkl}

\bibitem{MPI}
{M}essage {P}assing {I}nterface ({MPI}) forum (Aug 2017),
  \url{http://mpi-forum.org}

\bibitem{NvidiaProfiling}
Profiler :: {CUDA} toolkit documentation (Aug 2017),
  \url{http://docs.nvidia.com/cuda/profiler-users-guide/index.html}

\bibitem{Qt}
{Qt} | cross-platform software development for embedded \& desktop (Aug 2017),
  \url{https://www.qt.io/}

\bibitem{Shende:2011}
Shende, S., Malony, A.D., Spear, W., Schuchardt, K.: Characterizing {I/O}
  performance using the {TAU} performance system. In: PARCO. pp. 647--655
  (2011)

\bibitem{Vanderspoel:2005}
Van Der~Spoel, D., Lindahl, E., Hess, B., Groenhof, G., Mark, A.E., Berendsen,
  H.J.: {GROMACS}: fast, flexible, and free. Journal of computational chemistry
   26(16),  1701--1718 (2005)

\bibitem{Intelvtune}
{Intel}\textregistered ~{VTune}\texttrademark ~{Amplifier} (Aug 2017),
  \url{https://software.intel.com/en-us/intel-vtune-amplifier-xe}

\end{thebibliography}

\end{document}